\documentstyle[12pt,a4]{article}

\renewcommand{\a}{\alpha}
\renewcommand{\pb}{\overline{\partial}}
\newcommand{\p}{\partial}
\begin{document}

\title{\vspace{-2cm} \hfill {\small NTUA-45/94}\vspace{-.3cm} \\
\hfill {\small hep-th/9406136}\\ \vspace{-.3cm} \hfill {\small June
1994} \\ \vspace{1cm} {\Large \bf All WZW models in $D\leq5$}}
\vspace{2.5cm} \author{
{\normalsize \bf Alexandros A. Kehagias}
%\thanks{Partially supported by CEC Contract No. ERBCHBGCT920197}
 \thanks{e-mail: kehagias@sci.kun.nl}
\\ {\normalsize     Physics Department, } \\ {\normalsize National
Technical University,} \\ {\normalsize    15780 Zografou, Athens,
Greece
}\\ [.1cm]} \date{} \maketitle

\begin{abstract} \begin{sloppypar} \normalsize
We present here all the real algebras $\cal{A}$ with dim$\cal{A}\leq
$5 and all 6-dimensional nilpotent ones with symmetric,
invariant and non-degenerate  metrics for which a WZW model can be
constructed.
 In three and four dimensions there are no other algebras  than
the well known $SU(2)$, $SU(1,1)$, $E_2^c$ and $H_4$.
There exist only one five-dimensional  and one six-dimensional
nilpotent algebra with invariant non-degenerate metric and central
charge $c=5, \,6$, respectively.  We examine
in details the five-dimensional case  and, by  gauging an
 appropriate subgroup, four-dimensional plane-wave
 string backgrounds are obtained. The corresponding
background for the six-dimensional case is flat.

  \end{sloppypar} \end{abstract}

\newpage
\section{Introduction}
An important problem in string theory is to find
  exact string backgrounds. These will determine the short distance structure
of space-time and will provide informations about string-generated
gravitational
interactions.
 One may distinguishes roughly
the following classes of exact solutions, flat space with linear dilaton
\cite{1},  plane wave \cite{2} and
 $N=4$ supersymmetric string backgrounds
\cite{kkl}, WZW models \cite{6} and gauged WZW models \cite{4,4'}. We will
examine here solutions in the last two classes  writing down  all WZW models
based on groups with dimension up to five as well as on six-dimensional
nilpotent groups.

Conformal field theories can be constructed, in this or in the other
way, using
current algebras. The simplest models are the WZW models based on a
group G where the full symmetry of the action is realized in terms of
current algebras \cite{6}. Gauging appropriate (anomaly free) subgroups we
obtain new conformal field theories, the coset models \cite{halp}. WZW models
based on compact groups  have been employed in string
compactification while  those based on non-compact ones may regarded
as curved backgrounds \cite{5a}. Recently, attention has been given in
models based on non-semisimple groups \cite{8}--\cite{5'}.
 The non-invertibility of the Killing
form makes impossible in this case to construct the corresponding WZW
 models in the standard fashion. However, there exist groups which, although
non-semisimple, they have an invariant, symmetric and
non-degenerate metric. In this case, an affine Sugawara construction
can be carried out leading to integer central charge. Attempts to find
non-semisimple affine Sugawara constructions with non-integer central
charge \cite{11} show that in this case the construction factorizes into a
semisimple standard Sugawara construction and a non-semisimple one  with
integer central charge.

 The first of these WZW models was based on the
group $E_2^c$, a
central extension of the 2-dim Euclidean group and the corresponding
$\sigma$-model describes string
propagation on a four-dimensional gravitational plane wave background
\cite{8}.
This construction was subsequently extended to other non-semisimple
groups \cite{9}--\cite{12},\cite{14}, the representations of the affine
$E_2^c$ were constructed in terms of free fields and various gaugings
have been considered \cite{10,13,15}.
 The basic feature of all these models is that they describe
string propagation in backgrounds which admit one or more null Killing
vectors. They are  geodesically complete (free
of singularities) \cite{16} and, furthermore, it was shown that any
gravitational wave can be extended to an exact string background \cite{5} where
the underlying conformal field theory remains to be discovered \cite{5'}.

Here we examine algebras with invariant symmetric and non-degenerate
metric. Since all real algebras with dimension up to five and
nilpotent six dimensional ones are known, it is straightforward to
find those algebras for which such a metric exists.
 We found
that the first non-trivial case is the well-known three-dimensional
SU(2) and SU(1,1) and in four dimensions
the centrally extended Euclidean and the Heisenberg algebra. Finally,
there exists only one five-dimensional  and one six-dimensional
nilpotent algebra for which a WZW model can be constructed. We examine
in details the five-dimensional case  and, by  gauging an
 appropriate subgroup,
a four-dimensional string background is obtained. The corresponding
background for the six-dimensional case is flat.

\section{Invariants of Lie algebras}
A central problem in representation theory of Lie groups and Lie algebras
 is the determination of their invariants, i.e.,
 functions of the generators which commute with all generators.
{}From the mathematical point of view, their importance follows from
the fact that they  label representations,  split
reducible representations into irreducible ones,  produce all
special functions as eigenfunctions of the corresponding invariant
operators etc.. On the other hand, from the physical point of view,
the invariants of the symmetry group of a physical system provide
quantum numbers, mass formulae, energy spectra and so on.
 Let us stress, however, that  the invariants of a Lie group
are not necessarily Casimir operators (polynomials of the generators).
 There are groups with
polynomial invariants which give rise to Casimir operators,
as well as groups with rational invariants (ratio of polynomials) or even
groups with no invariants at all \cite{17,18}.

The invariants of semisimple groups are known and their number equals
the rank of the group.  For non-semisimple groups, there exists a
systematic way to compute their invariants.
The most well studied group in this category is  the Poincar\'e group
 the invariants of which
provide the  particle states with mass and spin.

An invariant of fundamental importance  is the quadratic
Casimir. It is quadratic in the  generators $J_i,\,  (i=1,2,...,dimG)$
 of the group G and can collectively be written as
\begin{equation}
C^{(2)}= \Omega^{ij} J_i J_j \,  ,\label{1}
\end{equation}
where $\Omega^{ij}$ may be viewed as elements of a symmetric and
possibly degenerate matrix.
The generators $J_i$ satisfy
\begin{equation}
[J_i,J_j]={f_{ij}}^k J_k \, ,\label{2}
\end{equation}
where ${f_{ij}}^k$ are the structure constants. From
 the invariance of $C^{(2)}$
it follows that
 $\Omega^{ij}$   must obey
\begin{equation}
\Omega^{ij}{f_{ki}}^\ell + \Omega^{i\ell}{f_{ki}}^j=0 \, . \label{3}
\end{equation}

If we consider $\Omega^{ij}$ as a symmetric matrix and if its determinant is
not zero, one can form the inverse matrix $\Omega_{ij}$ which
is also   symmetric,
 non-degenerate  and invariant under the adjoint action of the group
\begin{equation}
\Omega_{ij} {f_{k\ell}}^n + \Omega_{i\ell}{f_{kj}}^n=0 \, .\label{4}
\end{equation}
One  may then use $\Omega_{ij}$ as the metric on the group manifold and for
semisimple groups $\Omega_{ij}$ is proportional to the Killing form
$g_{ij}={f_{ik}}^\ell {f_{j\ell}}^k$.

 Non-semisimple groups may also have
quadratic Casimirs with non-degenerate $\Omega^{ij}$ although their Killing
form is degenerate. We recall
 the Poincar\'e
group in three dimensions which, apart from the usual
$mass^2=\vec{P}^{.}\vec{P}$ invariant, it has an additional one,
 the helicity defined as $\vec{J}^{.}\vec{P}$. As a result, the
Poincar\'e group in three dimensions is a six-dimensional non-semisimple
group with quadratic Casimir
\begin{equation}
C^{(2)}= k_1\vec{P}^{.}\vec{P} + k_2 \vec{J}^{.}\vec{P} \label{P}
\end{equation}
where $k_1\, , k_2$ are constants.
In this case, although the Killing form is degenerate the corresponding
$\Omega_{ij}$ is not and can be considered as the bi-invariant metric
on the group manifold. One may observe the existence of parameters
in $C^{(2)}$. The reason is that there are two linearly independent quadratic
Casimirs. Generally speaking,  the number $\tau$
 of independent Casimirs of an
algebra  $\cal A$ satisfies $\tau \leq dim{\cal A} - r({\cal A})$ where
$r(\cal A)$ is the rank of the matrix
\[
R_{ij}=[J_i,J_j].
\]
The equality holds for semisimple and nilpotent groups which have only
polynomial invariants and fails for an abstract algebra since
there exist in general and non-polynomial invariants \cite{19}.

 There are also exist other non-semisimple groups
with a non-degenerate metric, as the centrally extended Euclidean
group $E_d^c$ \cite{8}--\cite{10}, the Heisenberg group \cite{14} and so on.
 However, we do not
know all the groups with bi-invariant metric $\Omega_{ij}$, although
theorems for the structure of these groups have been proved \cite{20}.
 On the other
hand, all real Lie algebras
 with dimension less than five  $(dimG\leq5)$ \cite{21} and all nilpotent
six-dimensional ones \cite{22} are known.
Thus, in this case it is straightforward to  write down the algebras
with non-degenerate metric. We will follow the list presented in
\cite{18} where the independent Casimirs are given as well.

\vspace{.4cm}

{\bf $\alpha$) 1-dimensional real Lie algebras }
\vspace{.3cm}

The  one-dimensional algebras, denoted by $A_1$,
 are abelian and  the quadratic Casimir is just
the square of their generators.

\vspace{.4cm}

{\bf $\beta$) 2-dimensional real Lie algebras }

\vspace{.3cm}

There exist two two-dimensional Lie groups, the $A_{2,1}$ and the $A_{2,2}$.
 The former is    abelian $A_{2,1}=
A_1\oplus A_1$ with Casimir
\begin{equation}
C^{(2)} = J_1^2 + kJ_2^2 \, .
\end{equation}
The latter $A_{2,2}$ is solvable and it is defined by the commutation relation
\begin{equation}
[J_1,J_2]=J_1 \, . \label{5}
\end{equation}
This group is the  affine group in one dimension and
it has no invariants at all.

\vspace{.4cm}

{\bf $\gamma$) 3-dimensional real Lie algebras}
\vspace{.3cm}

The three-dimensional Lie algebras have been classified by Bianchi and
are known as Bianchi-type. There exist nine Bianchi-type
algebras (one of which depends on a parameter and it is
actually a continuous family)
 which together with the $A_1\!\oplus\!A_1\!\oplus\!A_1$ and
$A_{2,2}\!\oplus\!A_1$ consist the 11 different three-dimensional  real
Lie algebras. From these, only the Bianchi-type VII and IX  (denoted by
$A_{3,8}\, ,A_{3,9}$,  respectively,)
have quadratic Casimirs. These algebras are defined by the commutation
relations
\begin{eqnarray}
[J_1,J_2]=J_3 \, , &  [J_2,J_3]=J_1    \, , &  [J_3,J_1]=J_2 \, ,\nonumber  \\
\ [J_1,J_2]=-2J_3 \, , &  [J_2,J_3]=J_1 \, , &  [J_3,J_1]=J_2 \, ,
\end{eqnarray}
and the corresponding quadratic Casimirs are
\begin{eqnarray}
C^{(2)} & = & J_1^2+J_2^2+J_3^2 \nonumber \\
C^{(2)} & = & J_1J_3 +J_3J_1 +2J_2^2 \, . \label{6}
\end{eqnarray}

$A_{3,8}\, , A_{3,9}$ are semisimple and isomorphic
 to $SU(2), \, SL(2,R)$, respectively.

\vspace{.4cm}

{\bf $\delta$) 4-dimensional real Lie algebras}

\vspace{.3cm}

There exist 12 real four-dimensional Lie algebras. Five of them depend
on parameters and thus they form continuous families of algebras. There are
two algebras which have invariant metric, the solvable
algebras $A_{4,8}$
and $A_{4,10}$. The former is defined by the commutation relations
\begin{eqnarray}
[J_2,J_3] & = & J_1 \, , \nonumber \\
\ [J_2,J_4] & = & J_2 \, , \nonumber \\
\ [J_3,J_4] & = & -J_3\, . \label{9}
\end{eqnarray}
It is isomorphic to the real Heisenberg algebra $H(4,R)$.
The quadratic Casimir is
\begin{eqnarray}
C^{(2)}=k_1 J_1^2 + k_2(J_2J_3+J_3J_2-2J_1J_4) \, , \label{10}
\end{eqnarray}
and thus
 $\Omega^{ij}\, , \Omega_{ij}$ are given by
 \begin{equation}
\Omega^{ij} \;=\; \left( \begin{array}{cccc} k_1&0&0&-k_2 \\ 0&0&k_2&0
 \\ 0&k_2&0&0 \\ -k_2&0&0&0 \end{array} \right) \, , \,  \Omega_{ij} \;=\;
 \left( \begin{array}{cccc} 0&0&0&-q_2 \\ 0&0&q_2&0
 \\ 0&q_2&0&0 \\ -q_2&0&0&q_1 \end{array}
\right) \, ,
\label{11} \end{equation}
where $q_1=-k_1/k_2^2 \, , q_2=1/k_2$.

The algebra $A_{4,10}$ is defined by the commutation relations
\begin{eqnarray}
[J_2,J_3] & = & J_1 \, , \nonumber \\
\ [J_2,J_4] & = & -J_3 \, , \nonumber \\
\ [J_3,J_4] & = & J_2\, . \label{12}
\end{eqnarray}
and it is isomorphic to the centrally extended Euclidean algebra $E_2^c$.
The quadratic Casimir is
\begin{equation}
C^{(2)}=k_1J_1^2+k_2(J_2^2+J_3^2+2J_1J_4) \, , \label{13}
\end{equation}
and the metric is found to be
 \begin{equation}
\Omega^{ij} \;=\; \left( \begin{array}{cccc} k_1&0&0&k_2 \\ 0&k_2&0&0
 \\ 0&0&k_2&0 \\ k_2&0&0&0 \end{array} \right) \, , \,  \Omega_{ij} \;=\;
 \left( \begin{array}{cccc} 0&0&0&q_2 \\ 0&q_2&0&0
 \\ 0&0&q_2&0 \\ q_2&0&0&q_1 \end{array}
\right) \, ,
\label{14} \end{equation}
where $q_1=-k_1/k_2^2\, , q_2=1/k_2$.
The two algebras above are the only indecomposable
ones with  quadratic Casimirs. There also exist decomposable algebras
 with quadratic Casimirs, namely,  $A_1\!\oplus\!A_1\!\oplus\!A_1\!
\oplus\!A_1$
, $A_{3,8}\!\oplus\!A_1$ and $A_{3,9}\!\oplus\!A_1$. Note  that $A_{4,8}$
over the complex is isomorphic to $A_{4,10}$ (they have the same
complexification).

\vspace{.4cm}

{\bf $\epsilon$) 5-dimensional real Lie algebras}
\vspace{.3cm}

There exist 40 five-dimensional real indecomposable
 algebras eighteen of which depend on one
or more parameters and only one of these has an
 invariant metric. It is the nilpotent algebra $A_{5,3}$
which is defined by the commutation relations
\begin{eqnarray}
[J_3,J_4] & = & J_2 \, , \nonumber \\
\ [J_3,J_5] & = & J_1 \, , \nonumber \\
\ [J_4,J_5] & = & J_3\, . \label{15}
\end{eqnarray}
The quadratic Casimir for this algebra is
\begin{equation}
C^{(2)}=k_1J_1^2+k_2J_2^2+k_3(J_3^2+2J_2J_5-2J_1J_4)\, . \label{16}
\end{equation}
and the invariant metric is then found to be
\begin{equation}
\Omega^{ij} \;=\; \left( \begin{array}{ccccc} k_1&0&0&-k_3&0 \\
 0&k_2&0&0&k_3
 \\ 0&0&k_3&0&0 \\
-k_3&0&0&0&0 \\ 0&k_3&0&0&0  \end{array} \right) \, , \,  \Omega_{ij} \;=\;
 \left( \begin{array}{ccccc} 0&0&0&-q_1&0 \\ 0&0&0&0&q_1
 \\ 0&0&q_1&0&0 \\ -q_1&0&0&q_2&0 \\
0&q_1&0&0&q_3 \end{array}
\right) \, ,
\label{17} \end{equation}
where $q_1=1/k_3 \, , q_2=-k_1/k_3^2 \, , q_3=-k_2/k_3^2$.
The full list of real five-dimensional algebras
with invariant metric  contains also the decomposable algebras
$\oplus^5\!A_1$,  $A_{3,8}\!\oplus\!A_1\oplus\!A_1$,
$A_{3,9}\!\oplus\!A_1 \!\oplus\!A_1$,
$A_{4,8}\!\oplus\!A_1$ and $A_{4,10}\! \oplus\! A_1$.

\vspace{.4cm}

{\bf $\sigma\tau$) 6-dimensional nilpotent Lie algebras}

\vspace{.3cm}

There is no complete classification of the six-dimensional real algebras.
However, all nilpotent six-dimensional algebras are known. There exist
22 of them and only one has an invariant metric.
This is the algebra $A_{6,3}$  defined by the commutation relations
\begin{eqnarray}
[J_1,J_2] & = & J_6 \, , \nonumber \\
\ [J_1,J_3] & = & J_4 \, , \nonumber \\
\ [J_2,J_3] & = & J_5 \ . \label{18}
\end{eqnarray}
The quadratic Casimir is
\begin{equation}
C^{(2)}=k_1J_4^2+k_2J_5^2+k_3J_6^2+k_4(J_1J_5+J_3J_6-J_2J_4) \label{19}
\end{equation}
and the metric is given by
\begin{equation}
\Omega^{ij} \;=\; \left( \begin{array}{cccccc} 0&0&0&0&k_4&0\\
 0&0&0&-k_4&0&0
 \\ 0&0&0&0&0&k_4 \\
0&-k_4&0&k_1&0&0 \\ k_4&0&0&0&k_2&0 \\
0&0&k_4&0&0&k_3 \end{array} \right) \, , \,  \Omega_{ij} \;=\;
 \left( \begin{array}{cccccc} q_1&0&0&0&q_2&0 \\ 0&q_3&0&-q_2&0&0
 \\ 0&0&q_4&0&0&q_2 \\ 0&-q_2&0&0&0&0 \\
q_2&0&0&0&0&0 \\ 0&0&q_2&0&0&0 \end{array}
\right) \, ,
\label{20} \end{equation}
where $q_1=-k_2/k_4^2\, , q_2=1/k_4 \, , q_3=-k_1/k_4^2 \, , q_4=-k_3/k_4^2$.
The algebra $A_{6,3}$ together with $\oplus^6A_1$ and $A_{5,3}\oplus A_1$
completes the list of all real six-dimensional Lie algebras with an
invariant metric.

\section{WZW models}

A WZW model  based on a simple group G is defined on a two surface
$\Sigma$ by the action
 \begin{equation}
  {\cal S}_{WZW} \,=\, \frac{\kappa}{4 \pi} \int_{\Sigma}
d\sigma^2 Tr(g^{-1}\partial g \,
  g^{-1}\bar\partial g ) -i \frac{\kappa}{6 \pi} \int_{B}
d\sigma^3\epsilon^{abc} Tr(g^{-1}\partial_ag\wedge g^{-1}\partial_bg
  \wedge g^{-1}\partial_cg) \, , \label{21}
 \end{equation}
where $B$ is a three-manifold bounded by $\Sigma$.
If the traces are in some representation $\cal R$ of the group, then the
normalization factor $\kappa$ is given by
\[
\kappa=\frac{kc_Gd_G}{4\stackrel{\sim}{h}_Gc_Rd_R}
\]
where  k $\in $ Z is an integer, $d_R(d_G)$ and $c_R(c_G)$ are the dimension
and the quadratic Casimir of the $\cal R$ ($adj$) representation and
 $\stackrel{\sim}{h}_G$ is the
dual Coxeter number.

It is also possible to write down WZW models for non-semisimple groups
as well. The necessary condition is the existence of an (bi-)invariant
metric. This can be done as follows. One  defines on the group manifold
 the left and right invariant forms
  $g^{-1}dg$ and $dgg^{-1}$, respectively. Since these forms
are elements of the corresponding algebra, they may be  expressed as
\begin{eqnarray}
g^{-1}dg & =& L^i J_i = L_a^iJ_idz^a \, , \label{22} \\
dgg^{-1} & =& R^i J_i = R_a^iJ_idz^a \, , \label{23}
\end{eqnarray}
where $(z^a)$ parametrizes the Euclidean 2d world sheet. If the algebra
has an invariant metric $\Omega_{ij}$, then the corresponding
  WZW model can be written as
 \begin{equation}
{\cal S}_{WZW} \,=\, \frac{k}{4
\pi} \int_{\Sigma} d^{2} \sigma L^{i}_aL^{ja} \Omega_{ij}
\,\ + \frac{ik}{12 \pi} \int_{B} d^{3}\sigma \, \epsilon^{abc}
L^{i}_a L^{j}_b L^{k}_c {f_{ij}}^{\ell}
\Omega_{k\ell}\, .  \label{24}
 \end{equation}

The currents associated with the above action satisfy the current algebra
of G which is specified by the OPEs
\begin{equation}
J_i(z)J_j(w)=\frac{\Omega_{ij}}{(z-w)^2}+{f_{ij}}^k\frac{J_k(z)}{(z-w)}
+regular \, . \label{25}
\end{equation}
Starting with the current algebra one can construct stress tensors
which are bilinear in the currents, $T_L= L^{ij}
\stackrel{\textstyle .}{.}
J_iJ_j\stackrel{\textstyle .}{.}$ where $L^{ij}$ is the inverse
inertial tensor \cite{23}.

The condition for the $J_i$'s to be primary fields of weight 1 is written as
\begin{equation}
T_L(z)J_i = \frac{J_i(z)}{(z-w)^2}+\frac{\partial J_i(w)}{(z-w)}+
regular \, . \label{26}
\end{equation}
By using eqs.(\ref{25},\ref{26}) and the definition of $T_L$, we find that
$L^{ij}$ must satisfy \cite{12,23}
\begin{eqnarray}
   L^{ij} {f_{kj}}^{\ell} \,+\, L^{j\ell} {f_{kj}}^{i} \;&=&\; 0 \, ,
   \label{27} \\ 2 L^{ij}\Omega_{kj} + L^{mn}{f_{km}}^{\ell}
 {f_{\ell n}}^{i}\;&=&\; {\delta^{i}}_{k} \, . \label{28}
 \end{eqnarray}
Thus, $L^{ij}$ is an invariant symmetric tensor as follows from
eq.(\ref{27}) and by employing the latter in eq.(\ref{28}) we get
\begin{equation}
      L^{ij} ( 2 \Omega_{kj} + g_{kj} ) \;=\; {\delta^{i}}_{k}
      \label{29} \, , \end{equation} where $g_{kj}$
 is the Killing form.  Thus,
$L^{ij}$ is the inverse of the matrix $(2\Omega_{kj}+g_{kj})$ and
the central charge is
\begin{equation}
c=2L^{ij}\Omega_{ij} \, . \label{c}
\end{equation}
By using eqs(\ref{11},\ref{14},\ref{17},\ref{20}),
 we find that the $L^{ij}$ for the algebras
$A_{4,8}\, ,A_{4,10}\, , A_{5,3}$ and $A_{6,3}$ are
\begin{eqnarray}
L_{4,8}^{ij} \;=\;\frac{1}{2} \left( \begin{array}{cccc} k_2^2+k_1&0&0&-k_2
 \\ 0&0&k_2&0
 \\ 0&k_2&0&0 \\ -k_2&0&0&0 \end{array} \right) & , &
L_{4,10}^{ij} \;=\; \frac{1}{2}
\left( \begin{array}{cccc} k_2^2+k_1&0&0&k_2
 \\ 0&k_2&0&0
 \\ 0&0&k_2&0 \\ k_2&0&0&0 \end{array} \right) \, , \nonumber \\
L_{5,3}^{ij} \;=\;\frac{1}{2} \left( \begin{array}{ccccc} k_1&0&0&-k_3&0 \\
 0&k_2&0&0&k_3
 \\ 0&0&k_3&0&0 \\
-k_3&0&0&0&0 \\ 0&k_3&0&0&0  \end{array} \right)  & ,  &
L_{6,3}^{ij} \;=\; \frac{1}{2} \left( \begin{array}{cccccc} 0&0&0&0&k_4&0\\
 0&0&0&-k_4&0&0
 \\ 0&0&0&0&0&k_4 \\
0&-k_4&0&k_1&0&0 \\ k_4&0&0&0&k_2&0 \\
0&0&k_4&0&0&k_3 \end{array} \right) \, . \label{31}
\end{eqnarray}
 and the central charge is $c=4,\, 4,\, 5,\, 6$, respectively, i.e.
equals the dimension of the corresponding algebras \cite{11}.
The same result could be obtained by solving directly the Virasoro
master equation \cite{23}.

 One may read off a space-time and antisymmetric tensor field background
by identifying  the  WZW action with  the $\sigma$-model action
\begin{equation}
{\cal S} = \int d^2\sigma (G_{MN}\partial_a X^M\partial^aX^N
+iB_{MN}\epsilon^{ab}\partial_aX^M\partial_bX^M) \, . \label{32}
\end{equation}
We will examine only the cases of $A_{5,3}$ and $A_{6,2}$ since all the
rest have been studied so far.
To begin with the algebra $A_{5,3}$, we form the corresponding group
by exponentiation and we parametrize the group manifold with
coordinates $(\a_i, \, i=1,...,5)$ by writing the elements
of the group as
\begin{equation}
g=e^{\a_1J_1+\a_2J_2}e^{\a_3J_4}e^{\a_5J_5}e^{\a_4J_4} \, . \label{33}
\end{equation}
By using eqs(\ref{22},\ref{23}), we find that the left invariant
forms are
\begin{eqnarray}
L^1 & = & d\a_1+\frac{\a_5^2}{2}d\a_3 \, ,
\nonumber \\
L^2 & = & d\a_2+\a_4\a_5d\a_3-\frac{\a_4^2}{2}d\a_5\, ,
 \nonumber \\
L^3 & = & \a_5d\a_3-\a_4d\a_5 \, ,
 \nonumber \\
L^4 & = & d\a_3+d\a_4  \, , \nonumber \\
L^5 & = & d\a_5  \, , \label{l}
\end{eqnarray}
and the right invariant ones are
\begin{eqnarray}
  R^1 & = & d\a_1 +\frac{\a_5^2}{2}d\a_4\, ,
\nonumber \\
R^2 & = & d\a_2+\a_3\a_5d\a_4-\frac{\a_3^2}{2}d\a_5 \, , \nonumber \\
R^3 & = & \a_3d\a_5-\a_5d\a_4 \, , \nonumber \\
R^4 & = & d\a_3+d\a_4 \, , \nonumber \\
 R^5 & = & d\a_5 \, . \label{r}
\end{eqnarray}
Thus, the terms which are integrated
 in the action (\ref{24}) are calculated to be
\begin{eqnarray}
L_a^iL^{ja}\Omega_{ij} & = & 2q_1\partial_a\a_1\partial^a\a_3
-2q_1\partial_a\a_1\partial^a\a_4-q_1\a_5^2\partial_a\a_3\partial^a\a_4
\nonumber \\
& & +2q_1\partial_a\a_2\partial^a\a_5+q_2(\partial_a\a_3+\partial_a\a_4)^2
+q_3\partial_a\a_5\partial^a\a_5 \, ,
\label{37}
\end{eqnarray}
\begin{eqnarray}
\epsilon^{abc}L_a^iL_b^jL_c^k{f_{ij}}^\ell\Omega_{\ell k} & = &
6q_1\epsilon^{abc}\a_5\partial_a\a_3\partial_b\a_4\partial_c\a_5
\nonumber \\
& = & 3q_1\epsilon^{abc}\partial_c(\a_5^2\partial_a\a_3\partial_b\a_4)
\, , \label{38}
\end{eqnarray}
and the WZW action is written as
\begin{eqnarray}
{\cal S} &=&
  \frac{k}{4\pi} \int d^2\sigma \left(
2q_1\partial_a\a_1\partial^a\a_3
-2q_1\partial_a\a_1\partial^a\a_4-q_1\a_5^2\partial_a\a_3\partial^a\a_4
+2q_1\partial_a\a_2\partial^a\a_5+ \right. \nonumber \\ & & \left.
q_2(\partial_a\a_3+\partial_a\a_4)^2
 +q_3\partial_a\a_5\partial^a\a_5
+i q_1\epsilon^{ab}\partial_c\a_5^2\partial_a\a_3\partial_b\a_4\right )
\, . \label{39}
\end{eqnarray}
By identifying the WZW action above with the $\sigma$-model action
 (\ref{32})  ($X^M=(\a_1,...,\a_5)$), we may read off the space-time
metric and the antisymmetric tensor field
\begin{eqnarray}
  G_{ij} &=&
 \left( \begin{array}{ccccc} 0&0&-q_1&-q_1&0 \\ 0&0&0&0&q_1
 \\ -q_1&0&q_2&\frac{1}{2}(q_2-q_1\a_5^2)&0 \\
-q_1&0&\frac{1}{2}(q_2-q_1\a_5^2)&q_2&0 \\
0&q_1&0&0&q_3 \end{array}
\right) , \, \nonumber \\
B_{34}& =& \frac{\a_5^2}{2} \, .
\end{eqnarray}
The corresponding space-time line element and the antisymmetric field
$H_{MNL}=\partial_{[M}B_{NL]}$ are therefore
\begin{eqnarray}
ds^2 & = & -2q_1d\a_1d\a_3-2q_1d\a_1d\a_4-q_1\a_5^2d\a_3d\a_4
+q_2(d\a_3+d\a_4)^2
\nonumber \\
& & +2q_1d\a_2d\a_5+q_3d\a_5^2 \, , \label{42} \\
H_{345}& =& \a_5\, . \label{43}
\end{eqnarray}
We observe that the metric (\ref{42}) has signature $+1$ since it has two null
Killing vectors. Thus, it cannot be considered as a physical
space-time background, not even in a higher dimensional Kaluza-Klein
 setting. However, one may gauge an
anomaly-free subgroup in such a way to remove one time-like direction
\cite{4,4'}. This will be done in the next section.

In view of the conformal invariance of the WZW model, the one-loop
beta-function equations
\begin{equation}
R_{MN}-\frac{1}{4}H_{MN}^2-\nabla_M\nabla_N\phi =0 \, , \nonumber \\
\end{equation}
\begin{equation}
\nabla^M(e^{\phi}H_{MNL})  =0 \, , \nonumber \\
\end{equation}
\begin{equation}
-R+\frac{1}{12}H^2 +2\Delta\phi +\nabla_M\phi\nabla^M\phi
+\frac{2\delta c}{3}=0  \, , \label{w12}
\end{equation}
must be satisfied.  The shorthand
notation
 $H_{MN}^2={H_M}^{KL}H_{NKL}$,  $H^2=H_{MNL}H^{MNL}$
have been used and $\delta c$ is the central-charge deficit.  One can
verify that the metric (\ref{42}) is Ricci flat ($R_{MN}=0$),
$H_{MN}^2=0$ and thus, the beta-function  equations
are  indeed satisfied with $c=5$ and constant dilaton field.

Let us now turn into the six-dimensional algebra $A_{6,3}$. We parametrize
 the corresponding group with coordinates $(\a_i, \, i=1,\ldots,6)$ so
that its elements can be written as:
\begin{equation}
g=e^{\a_1J_1}e^{\a_2J_2}e^{\a_3J_3}e^{\a_4J_4+\a_5J_5+\a_6J_6}\, .
\label{47}
\end{equation}
By using eqs(\ref{22},\ref{23}), we find that the left invariant
forms are
\begin{eqnarray}
L^1 & = & d\a_1\, ,
\nonumber \\
L^2 & = & d\a_2\, ,
 \nonumber \\
L^3 & = & d\a_3 \, ,
 \nonumber \\
L^4 & = & d\a_4+\a_3d\a_1  \, , \nonumber \\
L^5 & = & d\a_5+\a_3d\a_2  \, , \nonumber \\
L^6 & = & d\a_6+\a_2d\a_1  \, , \label{l1}
\end{eqnarray}
and  the right invariant ones
\begin{eqnarray}
  R^1 & = & d\a_1 \, ,
\nonumber \\
R^2 & = & d\a_2 \, , \nonumber \\
R^3 & = & d\a_5 \, , \nonumber \\
R^4 & = & d\a_4+\a_1d\a_3 \, , \nonumber \\
R^5 & = & d\a_5+\a_2d\a_3 \, , \nonumber \\
 R^6 & = & d\a_6+\a_1d\a_2 \, . \label{r1}
\end{eqnarray}
Thus we have
\begin{eqnarray}
L_a^iL^{ja}\Omega_{ij} & = & q_1\partial_a\a_1\partial^a\a_1
+q_3\partial_a\a_2\partial^a\a_2
+q_4\partial_a\a_3\partial^a\a_3 \nonumber \\
& &
+2q_2\partial_a\a_1\partial^a\a_5
-2q_2\partial_a\a_2\partial^a\a_4+2q_2\partial_a\a_3\partial^a\a_6\nonumber \\
& &
+2q_2\a_2\partial_a\a_3\partial^a\a_1 \, ,
\label{50}
\end{eqnarray}
\begin{eqnarray}
\epsilon^{abc}L_a^iL_b^jL_c^k{f_{ij}}^\ell\Omega_{\ell k} & = &
6q_2\epsilon^{abc}\partial_a\a_1\partial_b\a_2\partial_c\a_3
\nonumber \\
& = & 6q_2\epsilon^{abc}\partial_a(\a_1\partial_b\a_2\partial_c\a_3)
\, . \label{51}
\end{eqnarray}
Substituting the above expressions in the WZW action and comparing with
the $\sigma$-model action (\ref{32}), the metric and the non-zero
components of the $H_{MNL}$-field
are given by
\begin{eqnarray}
ds^2& = & q_1d\a_1^2+q_3d\a_2^2+q_4d\a_3^2+2q_2d\a_1d\a_5-2q_2d\a_2d\a_4
\nonumber \\
& & +2q_2d\a_3d\a_6+2q_2\a_2d\a_3d\a_1 \, , \label{52} \\
H_{123}&=&q_2 \, .
\end{eqnarray}

One may easily verify that, by defining coordinates
\begin{eqnarray}
x_1 & = & \sqrt{q_1}\a_1+\frac{q_2}{\sqrt{q_1}}(\a_5+\frac{1}{2}\a_2\a_3)\, ,
 \nonumber \\
x_2 & = & \sqrt{q_3}\a_2-\frac{q_2}{\sqrt{q_3}}(\a_4+\frac{1}{2}\a_3\a_1)\, ,
 \nonumber \\
x_3 & = & \sqrt{q_4}\a_3+\frac{q_2}{\sqrt{q_4}}(\a_6+\frac{1}{2}\a_1\a_2)\, ,
 \nonumber \\
x_4 & = & \frac{q_2}{\sqrt{q_3}}(\a_4+\frac{1}{2}\a_1\a_3) \, , \nonumber \\
x_5 & = & \frac{q_2}{\sqrt{q_1}}(\a_5+\frac{1}{2}\a_2\a_3) \, , \nonumber \\
x_6 & = & \frac{q_2}{\sqrt{q_4}}(\a_6+\frac{1}{2}\a_1\a_2) \, ,
\end{eqnarray}
the metric takes the form
\begin{equation}
ds^2=dx_1^2+dx_2^2+dx_3^2-dx_4^2-dx_5^2-dx_6^2 \, ,
\end{equation}
while the non-vanishing components of the $H_{MNL}$-field are
\begin{eqnarray}
H_{123}=H_{162}=H_{134}=H_{164}=H_{532}=H_{526}=H_{543}=H_{564}=
\frac{q_2}{\sqrt{q_1q_2q_3}} \, .
\end{eqnarray}
Thus, the corresponding $\sigma$-model is flat with three time-like directions
and the beta-function equations are satisfied (with $c=6$)
in view of $H_{MN}^2=0$.

\section{Gauged WZW models}
The WZW action is invariant under global $g\rightarrow h_{L}^{-1}gh_R$
transformations. We can make this transformation local by
introducing gauge fields $A, \, \overline{A}$ with transformation
properties $A\rightarrow h_{L}^{-1}(A+\partial)h_L, \, \overline{A} \rightarrow
h_{R}^{-1}(\overline{A}+\pb)h_R$. The choice $h_R = h_L$
corresponds to vector gauging while $h_R=h_L^{-1}$ to axial one.
Anomaly considerations allow axial and vector gauging for abelian
subgroups and vector gauging for non-abelian ones. The gauged action
takes the form
\begin{equation}
S[g,A]=S[g]+\frac{k}{2\pi}\int d^2z(<A,\pb gg^{-1}>\mp
<\overline{A},g^{-1}\partial g>+A\overline{A} \mp
<g^{-1}Ag,\overline{A}> ) \, , \label{gw}
\end{equation}
where $<,>$ denotes the  inner product defined by $\Omega_{ij}$
 and upper and lower signs
correspond to vector and axial gauging, respectively \cite{4,4'}.

Let us now consider  axial gauging for the five-dimensional case
choosing to gauge  the abelian
subgroup generated by $J_4$. The gauged WZW action is written, in complex
coordinates, as
\begin{eqnarray}
S_{axial} & =&  \frac{k}{4\pi} \int d^{2}z
 \left\{ \right. -q_1\p\a_1\pb\a_3-q_1\p\a_3\pb\a_1-
q_1\p\a_1\pb\a_4-q_1\pb\a_1\p\a_4
   \nonumber \\
& &  -\frac{1}{2}q_1\a_5^2(\p\a_3\pb\a_4
+\p\a_4\pb\a_3)+ q_1(\p\a_5\pb\a_2+\p\a_2\pb\a_5)\nonumber \\
&& +q_2\p\a_3\pb\a_3
+q_2(\p\a_3\pb\a_4+\p\a_4\pb\a_3)+q_2\p\a_4\pb\a_4
\nonumber \\ && +\frac{1}{2}q_1
\a_5^2(\p\a_4\pb\a_3-\p\a_3\pb\a_4)
\nonumber \\ && +A\left[-2q_1(\pb\a_1+\frac{1}{2}
\a_5^2\pb\a_4)+2q_2(\pb\a_3+\pb\a_4)\right] \nonumber \\ &&
 +\bar{A}\left[ -2q_1(\p\a_1
+\frac{1}{2}\a_5^2\p\a_3)+2q_2(\p\a_3+\p\a_4)\right]
\nonumber \\ &&\left. +A\overline{A}(4q_2
-q_1\a_5^2)\right\} \label{a0}
\end{eqnarray}
and it is invariant under the transformations
\begin{eqnarray}
\delta\a_1=\delta\a_2=\delta\a_5=0 ,\nonumber \\
 \delta\a_3=\delta\a_4=\epsilon ,   \nonumber \\
\delta A=-\p\epsilon.
\end{eqnarray}
One can obtain the $\sigma$-model by fixing the gauge and then
integrate over the gauge field. A convenient gauge choice is $\a_3+
\a_4=0$ and the resulting $\sigma$-model action is
\begin{eqnarray}
S&=& \frac{k}{4\pi}\int
d^2z \left(
\right.\frac{4q_1q_2\a_5^2}{4q_2-q_1\a_5^2}\p\a\pb\a+q_1\p\a_5\pb\a_2
+q_1\pb\a_5\p\a_2+q_3\p\a_5\pb\a_5 \nonumber \\ &&
\left.-\frac{4q_1^2}{4q_2-q_1\a_5^2}
\p\a_1\pb\a_1-\frac{2q_1^2\a_5^2}{4q_2-q_1\a_5^2}(\pb\a\p\a_1-\p\a\pb\a_1)\right)
\, . \label{a} \end{eqnarray}

By comparing (\ref{a}) with the most general class of conformal
invariant $\sigma$-models
\begin{equation}
S=\int d^2z[(G_{\mu\nu}+B_{\mu\nu})\p X^{\mu}\pb X^{\nu}+
\a' R\Phi(X)] \, , \label{sigma}
\end{equation}
 we may read off the background
metric $G_{\mu\nu}$, the antisymmetric field $B_{\mu\nu}$ and the
dilaton $\Phi$ (coming from the $A\overline{A}$ term in the action
)
\begin{eqnarray}
ds^2 & =&
\frac{1}{u^2+\mu^2}dx^2+\frac{u^2}
 {u^2+\mu^2}dy^2+2dudv\, , \nonumber \\
B_{xy} & =& -\frac{1}{\mu}\frac{u^2}{u^2+\mu^2}, \nonumber \\
\Phi & = & ln(u^2+\mu^2) +const. \label{al3}.
\end{eqnarray}
We have defined $x=\sqrt{q_1}\a_1,\, y=\sqrt{q_2}\a, \, u=\a_5, \,
v=\frac{q_1}{4}\a_2 +\frac{q_3}{8}\a_5,\,  \mu^2=\frac{4q_2}{q_1}$ and
we have replaced $k$ by $(-k)$  and $q_3$ by $(-q_3)$ in the action (\ref{a}).
This plane-wave solution
has been found (with a different $B_{xy}$ but still the same $H_{uxy}$
 field) in \cite{5'} as a
special limit of the $E_2^c\times U(1)/U(1)$ coset model.

Let us now consider the vector gauging of the same abelian
subgroup generated by $J_4$. The gauged WZW action in complex
coordinates in this case is
\begin{eqnarray}
S_{vect} & = & \frac{1}{4\pi}\int d^2 z  \left\{ \right.  -q_1\p\a_1\pb\a_3
-q_1\p\a_3\pb\a_1-q_1\p\a_1\pb\a_4   \nonumber \\
& &  -\frac{1}{2}q_1\a_5^2(\p\a_3\pb\a_4
+\p\a_4\pb\a_3)+ q_1(\p\a_5\pb\a_2+\p\a_2\pb\a_5)\nonumber\\ &&
+q_2\p\a_3\pb\a_3
+q_2(\p\a_3\pb\a_4+\p\a_4\pb\a_3)+q_2\p\a_4\pb\a_4\nonumber \\ &&
+\p\a_5\pb\a_5+\frac{1}{2}q_1
\a_5^2(\p\a_4\pb\a_3-\p\a_3\pb\a_4) \nonumber \\
&&+A\left[-2q_1(\pb\a_1+\frac{1}{2}
\a_5^2\pb\a_4)+2q_2(\pb\a_3+\pb\a_4)\right]
\nonumber \\ && \left. -\overline{A}\left[ -2q_1(\p\a_1
+\frac{1}{2}\a_5^2\p\a_3)+2q_2(\p\a_3+\p\a_4)
\right]+q_1\a_5^2 A\overline{A}\right\}
\end{eqnarray}
This action is invariant under the transformations
\begin{eqnarray}
\delta\a_1=\delta\a_2=\delta\a_5=0 , \nonumber \\
 -\delta\a_3=\delta\a_4=\epsilon ,   \nonumber \\
\delta A=\p\epsilon.
\end{eqnarray}
 As in the case of axial gauging, by fixing the gauge and integrating
over the gauge field one can obtain the $\sigma$-model action which,
choosing $\a_3=\a_4(=\a)$ is found to be
\begin{eqnarray}
S&=&\frac{k}{4\pi}\int d^2 z  \left[ \right.
\frac{4q_2(4q_2-q_1\a_5^2)}{q_1\a_5^2}
\p\a\pb\a+\frac{4q_1}{\a_5^2}\p\a_1\pb\a_1+q_3\p\a_5\pb\a_5
\nonumber \\ & & \left. +q_1(\p\a_5\pb\a_2+\p\a_2\pb\a_5)
-\frac{8q_2}{\a_5^2}(\p\a\pb\a_1+\p\a_1\pb\a) \right]. \label{al3'}
\end{eqnarray}

By defining coordinates $x=\sqrt{q_1}(\a_1-\frac{2q_2}{q_1}\a),\,
y=\sqrt{q_2}\a, \, u=\a_5,
\, v=\frac{q_1}{4}\a_2+\frac{q_3}{8}\a_5$
 and comparing (\ref{al3'}) with (\ref{sigma}),
we may read off the space-time metric and the dilaton
\begin{eqnarray}
ds^2 & = &
\frac{1}{u^2}dx^2+dy^2+2dudv \, ,
\nonumber \\
\Phi & =& \ln u^2 +const. \label{al4}
\end{eqnarray}
One can verify that the one-loop
beta-function equations for conformal invariance are indeed satisfied
with central charge deficit $\delta c=0$.

Finally, let us stress that the solutions (\ref{al3},\ref{al4})
 are related to the
4-dim Minkowski space-time by a duality transformation \cite{24}. The duality
transformation for the case of an abelian isometry $x^0\rightarrow
x^0+const.$ in the coordinate system $(x^0,x^i)$ reads
\begin{eqnarray}
\tilde{G}_{00}&=&\frac{1}{G_{00}}\, , \,
\tilde{G}_{0i}=\frac{B_{i0}}{G_{00}} \, , \,  \tilde{G}_{ij}=G_{ij}-
\frac{G_{0i}G_{0j}-B_{0i}B_{0j}}{G_{00}} \nonumber \\
\tilde{B}_{0i}&=&\frac{G_{0i}}{G_{00}} \, , \,  \tilde{B}_{ij}=B_{ij}-
\frac{G_{0i}B_{0j}-G_{0j}B_{0i}}{G_{00}}\, , \nonumber \\
\tilde{\Phi}&=&\Phi+lnG_{00}.
\end{eqnarray}
Applying this transformation to the solution (\ref{al3}) for the
isometry $x\rightarrow x+const.$, and defining coordinates $x_1=\mu
x,\, x_2=x-\frac{y}{\mu}$, we get
\begin{eqnarray}
d\tilde{s}^2 & = & dx_1^2+u^2dx_2^2
+2dudv
 \, ,\nonumber \\
\tilde{B}_{xy}&=&0\, , \, \tilde{B}_{ij}=0 \nonumber \\
\tilde{\Phi}&=& const.
\end{eqnarray}
while the same isometry for the  solution (\ref{al4}) leads similarly
to
 \begin{eqnarray}
d\tilde{s}^2 & = & u^2dx^2+dy^2+2dudv
\, ,\nonumber \\
\tilde{\Phi}& =& const.
\end{eqnarray}
One may verify that the Riemann tensor for both the above metrics
vanishes and thus, they describe flat spacetimes.

\section{Conclusions}

Although many theorems about the structure of algebras with invariant
and non-degenerate metric exist, we do not know these algebras
explicity.
In particular, non-semisimple affine Sugawara constructions leads to
integer central charge and the corresponding exact background can be
considered as a physical space-time (as long as there exist one
time-like direction) for sting propagation. Thus, the non-semisimple
WZW models are particularly interesting since they provide exact
string backgrounds but, in view of the absence of any classification
of non-semisimple groups, their construction is an open problem.

Starting from the fact that
 all real algebras with dimension up to five and all
nilpotent six dimensional algebras are known, we examined for which of
these
an invariant, symmetric
and non-degenerate metric exists. We found that, up to four
dimensions,  there are no other
such algebras except the already known $SU(2)$, $SU(1,1)$, $E_2^c$ and
$H_4$. In five
dimensions there exist only one such algebra. We constructed the
corresponding WZW model and the resulting five-dimensional background
is of plane-wave type but with
two null Killing vectors. Thus, the resulting   space-time
cannot be considered as a physical background for string propagation.
However, by gauging  appropriate subgroups,  gauged WZW were
obtained corresponding to 4-dimensional plane-wave space-time
 with physical  Lorentz signature. These gauged models are
related to flat Minkowski space-time by a duality transformation. The
six dimensional case corresponds to flat space with three time-like
directions. \\

I would like to thank E. Angelopoulos for discussions, E. Kiritsis for comments
and N. Obers for an enlighting correspondance.

\end{document}